\documentclass{turing2012}
\usepackage{times}
\usepackage{amssymb, amsmath}
\usepackage[english]{babel}
\usepackage{cite}
\usepackage{colordvi}

\newcommand{\Eq}[1]{Eq.~(\ref{#1})}
\newcommand{\Sec}[1]{Sec.~\ref{#1}}
\newcommand{\pref}[1]{(\ref{#1})}

\newcommand{\folg}[3]{#1_{#2}, \ldots, #1_{#3}}
\newcommand{\D}{\, \mathrm{d}}
\newcommand{\vol}{\mathrm{vol}}

\renewcommand{\vec}[1]{\boldsymbol{#1}}

\begin{document}

\title{\bf Implementing Turing Machines \\
    in Dynamic Field Architectures\footnote{Appeared in M. Bishop und Y. J. Erden (Eds.) \emph{Proceedings of AISB12 World Congress 2012 --- Alan Turing 2012}, 5th AISB Symposium on Computing and Philosophy: Computing, Philosophy and the Question of Bio-Machine Hybrids, 36 -- 40.}
}

\author{Peter beim Graben\institute{
                                Institut f\"ur Deutsche Sprache und Linguistik,
                                Humboldt-Universit\"at zu Berlin,
                                email: peter.beim.graben@hu-berlin.de.}
    \and
    Roland Potthast\institute{
                                Dept. of Mathematics and Statistics,
                                University of Reading,
                                email: r.w.e.potthast@reading.ac.uk.}
}

\maketitle
\bibliographystyle{AISB}

\begin{abstract}
Cognitive computation, such as e.g. language processing, is conventionally regarded as Turing computation, and Turing machines can be uniquely implemented as nonlinear dynamical systems using generalized shifts and subsequent G\"odel encoding of the symbolic repertoire. The resulting nonlinear dynamical automata (NDA) are piecewise affine-linear maps acting on the unit square that is partitioned into rectangular domains. Iterating a single point, i.e. a microstate, by the dynamics yields a trajectory of, in principle, infinitely many points scattered through phase space. Therefore, the NDAs microstate dynamics does not necessarily terminate in contrast to its counterpart, the symbolic dynamics obtained from the rectangular partition. In order to regain the proper symbolic interpretation, one has to prepare ensembles of randomly distributed microstates with rectangular supports. Only the resulting macrostate evolution corresponds then to the original Turing machine computation. However, the introduction of random initial conditions into a deterministic dynamics is not really satisfactory. As a possible solution for this problem we suggest a change of perspective. Instead of looking at point dynamics in phase space, we consider functional dynamics of probability distributions functions (p.d.f.s) over phase space. This is generally described by a Frobenius-Perron integral transformation that can be regarded as a neural field equation over the unit square as feature space of a dynamic field theory (DFT). Solving the Frobenius-Perron equation, yields that uniform p.d.f.s with rectangular support are mapped onto uniform p.d.f.s with rectangular support, again. Thus, the symbolically meaningful NDA macrostate dynamics becomes represented by iterated function dynamics in DFT; hence we call the resulting representation \emph{dynamic field automata}.
\end{abstract}


\section{INTRODUCTION}
\label{sec:intro}

According to the central paradigm of classical cognitive science and to the Church-Turing thesis of computation theory (cf., e.g., \cite{Anderson95, HopcroftUllman79, Siegelmann95, Turing37}), cognitive processes are essentially rule-based manipulations of discrete symbols in discrete time that can be carried out by Turing machines. On the other hand, cognitive and computational neuroscience increasingly provide experimental and theoretical evidence, how cognitive processes might be implemented by neural networks in the brain.

The crucial question, how to bridge the gap, how to realize a Turing machine \cite{Turing37} by state and time continuous dynamical systems has been hotly debated by ``computationalists'' (such as Fodor and Pylyshyn \cite{FodorPylyshyn88}) and ``dynamicists'' (such as Smolensky \cite{Smolensky88a}) over the last decades. While computationalists argued that dynamical systems, such as neural networks, and symbolic architectures were either incompatible to each other, or the former were mere implementations of the latter, dynamicists have retorted that neural networks could be incompatible with symbolic architectures because the latter cannot be implementations of the former; see \cite{Graben04, Tabor09} for discussion.

Moore \cite{Moore90, Moore91a} has proven that a Turing machine can be mapped onto a generalized shift as a generalization of symbolic dynamics \cite{LindMarcus95}, which in turn becomes represented by a piecewise affine-linear map at the unit square using G\"odel encoding and symbologram reconstruction \cite{CvitanovicGunaratneProcaccia88, KennelBuhl03}. These \emph{nonlinear dynamical automata} have been studied and further developed by \cite{GrabenJurishEA04, GrabenGerthVasishth08}. Using a similar representation of the machine tape but a localist one of the machine's control states, Siegelmann and Sontag have proven that a Turing machine can be realized as a recurrent neural network with rational synaptic weights \cite{SiegelmannSontag95}. Along a different vain, deploying sequential cascaded networks, Pollack \cite{Pollack91} and later Moore \cite{Moore98} and Tabor \cite{Tabor00a, Tabor09} introduced and further generalized \emph{dynamical automata} as nonautonomous dynamical systems (see \cite{GrabenPotthast09a} for a unified treatment of these different approaches).

Inspired by population codes studied in neuroscience, Sch\"oner and co-workers devised \emph{dynamic field theory} as a framework for cognitive architectures and embodied cognition where symbolic representations correspond to regions in abstract feature spaces (e.g. the visual field, color space, limb angle spaces) \cite{ErlhagenSchoner02, SchonerDineva07}. Because dynamic field theory relies upon the same dynamical equations as \emph{neural field theory} investigated in theoretical neuroscience \cite{Amari77b, WilsonCowan73}, one often speaks also about \emph{dynamic neural fields} in this context.

In this communication we unify the abovementioned approaches. Starting from a nonlinear dynamical automaton as point dynamics in phase space in \Sec{sec:nda}, which bears interpretational peculiarities, we consider uniform probability distributions evolving in function space in \Sec{sec:dfa}. There we prove the central theorem of our proposal, that uniform distributions with rectangular support are mapped onto uniform distributions with rectangular support by the underlying NDA dynamics. Therefore, the corresponding dynamic field, implementing a Turing machine, shall be referred to as \emph{dynamic field automaton}. In the concluding \Sec{sec:discu} we discuss possible generalizations and advances of our approach. Additionally, we point out that symbolic computation in a dynamic field automaton can be interpreted in terms of contextual emergence \cite{AtmanspacherGraben07, AtmanspacherGraben09, BishopAtmanspacher06}.


\section{NONLINEAR DYNAMICAL AUTOMATA}
\label{sec:nda}

A nonlinear dynamical automaton (NDA: \cite{GrabenPotthast09a, GrabenGerthVasishth08, GrabenJurishEA04}) is a triple $M_{NDA} = (X, \mathcal{P}, \Phi)$ where $(X, \Phi)$ is a time-discrete dynamical system with phase space $X = [0, 1]^2 \subset \mathbb{R}^2$, the unit square, and flow $\Phi : X \to X$. $\mathcal{P} = \{ D_\nu | \nu = (i, j), 1 \le i \le m, 1 \le j \le n,  m, n \in \mathbb{N} \}$ is a rectangular partition of $X$ into pairwise disjoint sets, $D_\nu \cap D_\mu = \emptyset$ for $\nu \ne \mu$, covering the whole phase space $X = \bigcup_\nu D_\nu$, such that $D_\nu = I_i \times J_j$ with real intervals $I_i, J_j \subset [0, 1]$ for each bi-index $\nu = (i, j)$. Moreover, the cells $D_\nu$ are the domains of the branches of $\Phi$ which is a piecewise affine-linear map
\begin{equation}\label{eq:ndamap}
    \Phi(\vec{x}) =
    \begin{pmatrix}
      a^{\nu}_x \\
      a^{\nu}_y
    \end{pmatrix} +
       \begin{pmatrix}
      \lambda^{\nu}_x & 0 \\
      0 & \lambda^{\nu}_y
    \end{pmatrix} \cdot
    \begin{pmatrix}
      x \\
      y
    \end{pmatrix} \:,
\end{equation}
when $\vec{x} = (x, y)^T \in D_\nu$. The vectors $(a^{\nu}_x, a^{\nu}_y )^T \in \mathbb{R}^2$ characterize parallel translations, while the matrix coefficients $\lambda^{\nu}_x, \lambda^{\nu}_y \in \mathbb{R}_0^+$ mediate either stretchings ($\lambda > 1$), squeezings  ($\lambda < 1$), or identities ($\lambda = 1$) along the $x$- and $y$-axes, respectively.

The NDA's dynamics, obtained by iterating an orbit $\{ \vec{x}_t \in X | t \in \mathbb{N}_0 \}$ from initial condition $\vec{x}_0$ through
\begin{equation}
    \label{eq:ndadyn}
    \vec{x}_{t + 1} = \Phi(\vec{x}_t)
\end{equation}
describes a symbolic computation by means of a generalized shift \cite{Moore90, Moore91a} when subjected to the coarse-graining $\mathcal{P}$. To this end, one considers the set of bi-infinite, ``dotted'' symbolic sequences
\begin{equation}\label{eq:symseq}
    s = \ldots a_{i_{-3}} a_{i_{-2}} a_{i_{-1}} . a_{i_{0}} a_{i_{1}} a_{i_{2}} \ldots
\end{equation}
with symbols $a_{i_k} \in \mathbf{A}$ taken from a finite set, an alphabet $\mathbf{A}$. In \Eq{eq:symseq} the dot denotes the observation time $t = 0$ such that the symbol right to the dot, $a_{i_{0}}$, displays the current state, dissecting the string $s$ into two one-sided infinite strings $s = (s'_L, s_R)$ with $s'_L = a_{i_{-1}} a_{i_{-2}} a_{i_{-3}} \ldots$ as the left-hand part in reversed order and $s_R = a_{i_{0}} a_{i_{1}} a_{i_{2}} \ldots$ as the right-hand part. Applying a G\"odel encoding
\begin{eqnarray}
    \label{eq:goedel}
    x &=& \psi(s'_L) = \sum_{k = 1}^\infty \psi(a_{i_{-k}}) b_L^{-k} \\
    y &=& \psi(s_R) = \sum_{k = 0}^\infty \psi(a_{i_k}) b_R^{-k - 1} \nonumber
\end{eqnarray}
to the pair $s = (s'_L, s_R)$, where $\psi(a_j) \in \mathbb{N}_0$ is an integer G\"odel number for symbol $a_j \in \mathbf{A}$ and $b_L, b_R \in \mathbb{N}$ are the numbers of symbols that could appear either in $s_L$ or in $s_R$, respectively, yields the so-called symbol plane or symbologram representation $(x, y)^T$ of $s$ in the unit square $X$ \cite{CvitanovicGunaratneProcaccia88, KennelBuhl03}.

A generalized shift emulating a Turing machine\footnote{
    A generalized shift becomes a Turing machine by interpreting $a_{i_{-1}}$ as the current tape symbol underneath the head and $a_{i_{0}}$ as the current control state $q$. Then the remainder of $s_L$ is the tape left to the head and the remainder of $s_R$ is the tape right to the head. The DoD is the word $w = a_{i_{-1}} . a_{i_{0}}$ of length $d = 2$.
}
is a pair $M_{GS} = (\mathbf{A}^\mathbb{Z}, \Psi)$ where $\mathbf{A}^\mathbb{Z}$ is the space of bi-infinite, dotted sequences with $s \in \mathbf{A}^\mathbb{Z}$ and $\Psi : \mathbf{A}^\mathbb{Z} \to \mathbf{A}^\mathbb{Z}$ is given as
\begin{equation}\label{eq:genshift1}
    \Psi(s) = \sigma^{F(s)}(s \oplus G(s))
\end{equation}
with
\begin{eqnarray}
\label{eq:genshift}
  \label{eq:genshift2} F: \mathbf{A}^\mathbb{Z} &\to& \mathbb{Z} \\
  \label{eq:genshift3} G: \mathbf{A}^\mathbb{Z} &\to& \mathbf{A}^e \:,
\end{eqnarray}
where $\sigma : \mathbf{A}^\mathbb{Z} \to \mathbf{A}^\mathbb{Z}$ is the usual left-shift from symbolic dynamics \cite{LindMarcus95}, $F(s) = l$ dictates a number of shifts to the right ($l < 0$), to the left ($l > 0$) or no shift at all ($l = 0$), $G(s)$ is a word $w'$ of length $e \in \mathbb{N}$ in the domain of effect (DoE) replacing the content $w \in \mathbf{A}^d$, which is a word of length $d \in \mathbb{N}$, in the domain of dependence (DoD) of $s$, and $s \oplus G(s)$ denotes this replacement function.

From a generalized shift $M_{GS}$ with DoD of length $d$ an NDA $M_{NDA}$ can be constructed as follows: In the G\"odel encoding \pref{eq:goedel} the word contained in the DoD at the left-hand-side of the dot, partitions the $x$-axis of the symbologram into intervals $I_i$, while the word contained in the DoD at the right-hand-side of the dot partitions its $y$-axis into intervals $J_j$,  such that the rectangle $D_\nu = I_i \times J_j$ ($\nu = (i, j)$) becomes the image of the DoD. Moore \cite{Moore90, Moore91a} has proven that the map $\Psi$ is then represented by a piecewise affine-linear (yet, globally nonlinear) map $\Phi$ with branches at $D_\nu$.

In general, a Turing machine has a distinguished blank symbol, $\sqcup$ delimiting the machine tape and also some distinguished final states indicating termination of a computation \cite{HopcroftUllman79}. If there are no final states, the automaton is said to terminate with empty tape $s = \sqcup^\infty . \sqcup^\infty$. By mapping $\psi(\sqcup) = 0$ through the G\"odel encoding, the terminating state becomes a fixed point attractor $(0, 0)^T \in X$ in the symbologram representation. Moreover, sequences of finite length are then described by pairs of rational numbers by virtue of \Eq{eq:goedel}. Therefore, NDA Turing machine computation becomes essentially rational dynamics.

In the framework of generalized shifts and nonlinear dynamical automata, however, another solution appears to be more appropriate for at least three important reasons: Firstly, Siegelmann \cite{Siegelmann96b} further generalized generalized shifts to so-called analog shifts, where the DoE $e$ in \Eq{eq:genshift3} could be infinity (e.g. by replacing the finite word $w$ in the DoD by the infinite binary representation of $\pi$). Secondly, the NDA representation of a generalized shift should preserve structural relationships of the symbolic description, such as the word semigroup property of strings. Beim Graben et al. \cite{GrabenJurishEA04} have shown that a representation of finite strings by means of equivalence classes of infinite strings, the so-called cylinder sets in symbolic dynamics \cite{McMillan53} lead to monoid homomorphisms from symbolic sequences to the symbologram representation. Then, the empty word $\varepsilon$, the neutral element of the word semigroup, is represented by the unit interval $[0, 1]$ of real numbers. And thirdly, beim Graben et al. \cite{GrabenGerthVasishth08} combined NDAs with dynamical recognizers \cite{Pollack91, Moore98, Tabor00a} to describe interactive computing where symbols from an information stream were represented as operators on the symbologram phase space of an NDA. There, a similar semigroup representation theorem holds.

For these reasons, we briefly recapitulate the cylinder set approach here. In symbolic dynamics, a cylinder set is a subset of the space $\mathbf{A}^\mathbb{Z}$ of bi-infinite sequences from an alphabet $\mathbf{A}$ that agree in a particular building block of length $n \in \mathbb{N}$ from a particular instance of time $t \in \mathbb{Z}$, i.e.
\begin{eqnarray}\label{eq:cylinder}
    C(n, t) &=& [\folg a {i_1} {i_n} ]_t \nonumber \\
    &=& \{ s \in \mathbf{A}^{\mathbb{Z}} \,| \, s_{t + k - 1} = a_{i_k} , \quad k = 1, \dots, n \}
\end{eqnarray}
is called $n$-cylinder at time $t$. When now $t < 0, n > |t| + 1$ the cylinder contains the dotted word $w = s_{-1} . s_0$ and can therefore be decomposed into a pair of cylinders $(C'(|t|, t), C(|t| + n - 1, 0))$ where $C'$ denotes reversed order of the defining strings again. In the G\"odel encoding \pref{eq:goedel} each cylinder has a lower and an upper bound, given by the G\"odel numbers 0 and $b_L - 1$, $b_R - 1$, respectively. Then
\begin{eqnarray*}
  \inf(\psi(C'(|t|, t))) &=& \psi(\folg a {i_{|t|}} {i_1}) \\
  \sup(\psi(C'(|t|, t))) &=& \psi(\folg a {i_{|t|}} {i_1}) + b_L^{-|t|} \\
  \inf(\psi(C(|t| + n - 1, 0))) &=& \psi(\folg a {i_{|t| + 1}} {i_n}) \\
  \sup(\psi(C(|t| + n - 1, 0))) &=& \psi(\folg a {i_{|t| + 1}} {i_n}) + b_R^{-|t| - n + 1} \:,
\end{eqnarray*}
where the suprema have been evaluated by means of geometric series. Thereby, each part cylinder $C$ is mapped onto a real interval $[\inf(C), \sup(C)] \subset [0, 1]$ and the complete cylinder $C(n, t)$ onto the Cartesian product of intervals $R = I \times J \subset [0, 1]^2$, i.e. onto a rectangle in unit square. In particular, the empty cylinder, corresponding to the empty tape $\varepsilon . \varepsilon$ is represented by the complete phase space $X = [0, 1]^2$.

Fixing the prefixes of both part cylinders and allowing for random symbolic continuation beyond the defining building blocks, results in a cloud of randomly scattered points across a rectangle $R$ in the symbologram. These rectangles are consistent with the symbol processing dynamics of the NDA, while individual points $\vec{x} \in [0, 1]^2$ no longer have an immediate symbolic interpretation. Therefore, we refer to arbitrary rectangles $R \in [0, 1]^2$ as to NDA macrostates, distinguishing them from NDA microstates $\vec{x}$ of the underlying dynamical system. In other words, the symbolically meaningful macrostates are emergent on the microscopic NDA dynamics. We discuss in \Sec{sec:discu} how a particular concept, called contextual emergence, could describe this phenomenon \cite{AtmanspacherGraben07, AtmanspacherGraben09, BishopAtmanspacher06}.


\section{DYNAMIC FIELD AUTOMATA}
\label{sec:dfa}

From a conceptional point of view it does not seem very satisfactory to include such a kind of stochasticity into a deterministic dynamical system. However, as we shall demonstrate in this section, this apparent defect could be easily remedied by a change of perspective. Instead of iterating clouds of randomly prepared initial conditions according to a deterministic dynamics, one could also study the deterministic dynamics of probability measures over phase space. At this higher level of description, introduced by Koopman et al. \cite{Koopman31, KoopmanVonNeumann32} into theoretical physics, the point dynamics in phase space is replaced by functional dynamics in Banach or Hilbert spaces. This approach has its counterpart in neural \cite{Amari77b, WilsonCowan73} and dynamic field theory \cite{ErlhagenSchoner02, SchonerDineva07} in theoretical neuroscience.

In dynamical system theory the abovementioned approach is derived from the conservation of probability as expressed by a Frobenius-Perron equation \cite{Ott93}
\begin{equation}\label{eq:froper}
    \rho(\vec{x}, t) = \int_X \delta(\vec{x} - \Phi^{t - t'}(\vec{x}')) \rho(\vec{x}', t') \D \vec{x}' \:,
\end{equation}
where $\rho(\vec{x}, t)$ denotes a probability density function over the phase space $X$ at time $t$ of a dynamical system,  $\Phi^t : X \to X$ refers to either a continuous-time ($t \in \mathbb{R}_0^+$) or discrete-time ($t \in \mathbb{N}_0$) flow and the integral over the delta function expresses the probability summation of alternative trajectories all leading into the same state $\vec{x}$ at time $t$.


\subsection{Temporal Evolution}
\label{sec:evolv}

In the case of an NDA, the flow is discrete and piecewise affine-linear on the domains $D_\nu$ as given by \Eq{eq:ndamap}. As initial probability distribution densities $\rho(\vec{x}, 0)$ we consider uniform distributions with rectangular support $R_0 \subset X$, corresponding to an initial NDA macrostate,
\begin{equation}\label{eq:iniuni}
    u(\vec{x}, 0) = \frac{1}{|R_0|} \chi_{R_0}(\vec{x}) \:,
\end{equation}
where $|R_0| = \vol(R_0)$ is the ``volume'' (actually the area) of $R_0$ and
\begin{equation}\label{eq:charfn}
   \chi_{A}(\vec{x}) = \begin{cases}
                                        0 & \quad:\quad \vec{x} \notin A \\
                                        1 & \quad:\quad \vec{x} \in A
                                    \end{cases}
\end{equation}
is the characteristic function for a set $A \subset X$. A crucial requirement for these distributions is that they must be consistent with the partition $\mathcal{P}$ of the NDA, i.e. there must be a bi-index $\nu = (i, j)$ such that the support $R_0 \subset D_\nu$.

Inserting \pref{eq:iniuni} into the Frobenius-Perron equation \pref{eq:froper} yields for one iteration \begin{equation}\label{eq:froper2}
    u(\vec{x}, t + 1) = \int_X \delta(\vec{x} - \Phi(\vec{x}')) u(\vec{x}', t) \D \vec{x}' \:.
\end{equation}

In order to evaluate \pref{eq:froper2}, we first use the product decomposition of the involved functions:
\begin{equation}\label{eq:decouni1}
    u(\vec{x}, 0) = u_x(x, 0) u_y(y, 0)
\end{equation}
with
\begin{eqnarray}
    \label{eq:decounix}
    u_x(x, 0) &=& \frac{1}{|I_0|} \chi_{I_0}(x) \\
    \label{eq:decouniy}
    u_y(y, 0) &=& \frac{1}{|J_0|} \chi_{J_0}(y)
\end{eqnarray}
and
\begin{equation}\label{eq:decdelta}
  \delta(\vec{x} - \Phi(\vec{x}')) = \delta(x - \Phi_x(\vec{x}')) \delta(y - \Phi_y(\vec{x}'))  \:,
\end{equation}
where the intervals $I_0, J_0$ are the projections of $R_0$ onto $x$- and $y$-axes, respectively. Correspondingly, $\Phi_x$ and $\Phi_y$ are the projections of $\Phi$ onto $x$- and $y$-axes, respectively. These are obtained from \pref{eq:ndamap} as
\begin{eqnarray}
    \label{eq:mapx}
    \Phi_x(\vec{x}') &=& a^{\nu}_x + \lambda^{\nu}_x x' \\
    \label{eq:mapy}
    \Phi_y(\vec{x}') &=& a^{\nu}_y + \lambda^{\nu}_y y' \:.
\end{eqnarray}
Using this factorization, the Frobenius-Perron equation \pref{eq:froper2} separates into
\begin{eqnarray}
    \label{eq:fpx}
    u_x(x, t + 1) &=& \int_{[0, 1]} \delta(x - a^{\nu}_x - \lambda^{\nu}_x x') u_x(x', t) \D x' \\
    \label{eq:fpy}
    u_y(y, t + 1) &=& \int_{[0, 1]} \delta(y - a^{\nu}_y - \lambda^{\nu}_y y') u_y(y', t) \D y'
\end{eqnarray}

Next, we evaluate the delta functions according to the well-known lemma
\begin{equation}\label{eq:evadelta}
 \delta(f(x)) = \sum_{l : \text{simple zeros}} |f'(x_l)|^{-1} \delta(x - x_l) \:,
\end{equation}
where $f'(x_l)$ indicates the first derivative of $f$ in $x_l$. \Eq{eq:evadelta} yields for the $x$-axis
\begin{equation}\label{eq:zeros}
    x_\nu = \frac{x - a^{\nu}_x}{\lambda^{\nu}_x} \:,
\end{equation}
i.e. one zero for each $\nu$-branch, and hence
\begin{equation}\label{eq:slope}
    |f'(x_\nu')| = \lambda^{\nu}_x \:.
\end{equation}
Inserting \pref{eq:evadelta}, \pref{eq:zeros} and \pref{eq:slope} into \pref{eq:fpx}, gives
\begin{eqnarray*}
   u_x(x, t + 1) &=& \sum_\nu \int_{[0, 1]} \frac{1}{\lambda^{\nu}_x} \delta\left( x' - \frac{x - a^{\nu}_x}{\lambda^{\nu}_x} \right) u_x(x', t) \D x' \\
    &=& \sum_\nu \frac{1}{\lambda^{\nu}_x} u_x\left( \frac{x - a^{\nu}_x}{\lambda^{\nu}_x}, t \right)
\end{eqnarray*}

Next, we take into account that the distributions must be consistent with the NDA's partition. Therefore, for given $\vec{x} \in D_\nu$ there is only one branch of $\Phi$ contributing a simple zero to the sum above. Hence,
\begin{equation}\label{eq:uniter}
    u_x(x, t + 1) =
    \sum_\nu \frac{1}{\lambda^{\nu}_x} u_x\left( \frac{x - a^{\nu}_x}{\lambda^{\nu}_x}, t \right) =
    \frac{1}{\lambda^{\nu}_x} u_x\left( \frac{x - a^{\nu}_x}{\lambda^{\nu}_x}, t \right) \:.
\end{equation}

\begin{theorem}
The evolution of uniform p.d.f.s with rectangular support according to the NDA dynamics \Eq{eq:froper2} is governed by
\begin{equation}\label{eq:fpuniform}
    u(\vec{x}, t) = \frac{1}{|\Phi^t(R_0)|} \chi_{\Phi^t(R_0)}(\vec{x}) \:.
\end{equation}
\end{theorem}

\noindent\textbf{Proof (by means of induction).}

\noindent
1. Inserting the initial uniform density distribution \pref{eq:iniuni} for $t = 0$ into \Eq{eq:uniter}, we obtain by virtue of \pref{eq:decounix}
\[
        u_x(x, 1) = \frac{1}{\lambda^{\nu}_x} u_x\left( \frac{x - a^{\nu}_x}{\lambda^{\nu}_x}, 0 \right) =
        \frac{1}{\lambda^{\nu}_x} \frac{1}{|I_0|} \chi_{I_0}\left( \frac{x - a^{\nu}_x}{\lambda^{\nu}_x} \right) \:.
\]
Deploying \pref{eq:charfn} yields
\[
    \chi_{I_0}\left( \frac{x - a^{\nu}_x}{\lambda^{\nu}_x} \right) =
    \begin{cases}
        0 & \quad:\quad \frac{x - a^{\nu}_x}{\lambda^{\nu}_x}  \notin I_0 \\
        1 & \quad:\quad \frac{x - a^{\nu}_x}{\lambda^{\nu}_x} \in I_0 \:.
    \end{cases}
\]
Let now $I_0 = [p_0, q_0] \subset [0, 1]$ we get
\begin{eqnarray*}
    && \frac{x - a^{\nu}_x}{\lambda^{\nu}_x} \in I_0 \\
    &\Longleftrightarrow&
    p_0 \le \frac{x - a^{\nu}_x}{\lambda^{\nu}_x} \le q_0 \\
    &\Longleftrightarrow&
    \lambda^{\nu}_x p_0 \le x - a^{\nu}_x \le \lambda^{\nu}_x q_0 \\
    &\Longleftrightarrow&
    a^{\nu}_x + \lambda^{\nu}_x p_0 \le x \le a^{\nu}_x + \lambda^{\nu}_x q_0 \\
    &\Longleftrightarrow&
    \Phi_x(p_0) \le x  \le \Phi_x(q_0) \\
    &\Longleftrightarrow&
    x  \in \Phi_x(I_0) \:,
\end{eqnarray*}
Where we made use of \pref{eq:mapx}.

Moreover, we have
\[
 \lambda^{\nu}_x |I_0| = \lambda^{\nu}_x (q_0 - p_0) = q_1 - p_1 = |I_1|
\]
with $I_1 = [p_1, q_1] = \Phi_x(I_0)$. Therefore,
\[
        u_x(x, 1) = \frac{1}{|I_1|} \chi_{I_1}(x) \:.
\]

The same argumentation applies to the $y$-axis, such that we eventually obtain
\begin{equation}\label{eq:uniter2}
    u(\vec{x}, 1) = \frac{1}{|R_1|} \chi_{R_1}(\vec{x}) \:,
\end{equation}
with $R_1 = \Phi(R_0) $ the image of the initial rectangle $R_0 \subset X$. Thus, the image of a uniform density function with rectangular support is a uniform density function with rectangular support again.

2. Assume \pref{eq:fpuniform} is valid for some $t \in \mathbb{N}$. Then it is obvious that \pref{eq:fpuniform} also holds for $t + 1$ by inserting the $x$-projection of \pref{eq:fpuniform} into \pref{eq:uniter} using \pref{eq:decounix}, again. Then, the same calculation as under 1. applies when every occurrence of $0$ is replaced by $t$ and every occurrence of $1$ is replaced by $t + 1$.

By means of this construction we have implemented an NDA by a dynamically evolving field. Therefore, we call this representation \emph{dynamic field automaton (DFA)}.


\subsection{Kernel Construction}
\label{sec:kern}

The Frobenius-Perron equation \pref{eq:froper2} can be regarded as a time-discretized Amari dynamic neural field equation \cite{Amari77b} which is generally written as
\begin{equation}\label{eq:Amari}
  \tau \frac{\partial u(\vec{x}, t)}{\partial t} + u(\vec{x}, t) =
  \int_X w(\vec{x}, \vec{x}') f(u(\vec{x}', t)) \; \D \vec{x}' \:.
\end{equation}
Here, $\tau$ is the characteristic time constant of activation decay, $w(\vec{x}, \vec{x}')$ denotes the synaptic weight kernel, describing the connectivity between sites $\vec{x}, \vec{x}' \in X$ and $f$ is a typically sigmoidal activation function for converting membrane potential $u(\vec{x}, t)$ into spike rate $f(u(\vec{x}, t))$.

Discretizing time according to Euler's rule with increment $\Delta t = \tau$ yields
\begin{eqnarray*}
    \tau \frac{u(\vec{x}, t + \tau) - u(\vec{x}, t)}{\tau} + u(\vec{x}, t) &=&
  \int_X w(\vec{x}, \vec{x}') f(u(\vec{x}', t)) \; \D \vec{x}' \\
  u(\vec{x}, t + \tau) &=& \int_X w(\vec{x}, \vec{x}') f(u(\vec{x}', t)) \; \D \vec{x}' \:.
\end{eqnarray*}
For $\tau = 1$ and $f(u) = u$ the Amari equation becomes the Frobenius-Perron equation \pref{eq:froper2} when we set
\begin{equation}\label{eq:nftkernel}
    w(\vec{x}, \vec{x}') = \delta(\vec{x} - \Phi(\vec{x}')) \:.
\end{equation}
This is the general solution of the kernel construction problem \cite{PotthastGraben09a, GrabenPotthast09a}. Note that $\Phi$ is not injective, i.e.\ for fixed $x$ the kernel is a sum of delta functions coding the influence from different parts of the space $X = [0,1]^2$. Note further that higher-order discretization methods of explicit or implicit type such as the Runge-Kutta scheme could be applied to \Eq{eq:Amari} as well. But in this case the relationship between the Turing dynamics as expressed by the Frobenius-Perron equation \pref{eq:froper} and the neural field dynamics would become much more involved. We leave this as an interesting question for further research.


\section{DISCUSSION}
\label{sec:discu}

In this communication we combined nonlinear dynamical automata as implementations of Turing machines by nonlinear dynamical systems with dynamic field theory, where computations are characterized as evolution in function spaces over abstract feature spaces. Choosing the unit square of NDAs as feature space we demonstrated that Turing computation becomes represented as dynamics in the space of uniform probability density functions with rectangular support.

The suggested framework of dynamic field automata may exhibit several advantages. First of all, massively parallel computation could become possible by extending the space of admissible p.d.f.s. By allowing either for supports that overlap the partition of the underlying NDA or for multimodal distribution functions, one could prepare as many symbolic representations one wants and process them in parallel by the DFA. Moreover, DFAs could be easily integrated into wider dynamic field architectures for object recognition or movement preparation. They could be programmed for problem-solving, logical interferences or syntactic language processing. In particular, Bayesian inference or the processing of stochastic grammars could be implemented by means of appropriate p.d.f.s.

For those applications, DFAs should be embedded into time-continuous dynamics. This involves the construction of more complicated kernels through solving inverse problems along the lines of Potthast et al. \cite{PotthastGraben09a, GrabenPotthast09a}. We shall leave these questions for future research.

The construction of DFAs has also interesting philosophical implications. One of the long-standing problems in philosophy of science was the precise relationship between point mechanics, statistical mechanics and thermodynamics in theoretical physics: Is thermodynamics merely reducible to point mechanics via statistical mechanics? Or are thermodynamic properties such as temperature emergent on mechanical descriptions?

Due to the accurate analysis of Bishop and Atmanspacher \cite{BishopAtmanspacher06}, point mechanics and statistical mechanics simply provide two different levels of description: On one hand, point mechanics deals with the dynamics of microstates in phase space. On the other hand, statistical mechanics, in the formulation of Koopman et al. \cite{Koopman31, KoopmanVonNeumann32} (see \Sec{sec:dfa}), deals with the evolution of probability distributions over phase space, namely macrostates, in abstract function spaces. Both are completely disparate descriptions, none reducible to the other. However, the huge space of (largely unphysical) macrostates must be restricted to a subspace of physically meaningful thermal equilibrium states that obey a particular stability criterium (essentially the maximum-entropy principle). This restriction of states bears upon a contingent context, and in this sense, thermodynamic properties have been called \emph{contextually emergent} by \cite{BishopAtmanspacher06}.

Our construction of DFAs exhibits an interesting analogy to the relationship between mechanical micro- and thermal macrostates: Starting from microscopic nonlinear dynamics of an NDA, we used the Frobenius-Perron equation for probability density functions in order to derive an evolution law of macrostates: The time-discretized Amari equation \pref{eq:Amari} with kernel \pref{eq:nftkernel}. However, with respect to the underlying NDA, not every p.d.f. can be interpreted as a symbolic representation of a Turing machine configuration. Therefore, we had to restrict the space of all possible p.d.f.s, by taking only uniform p.d.f.s with rectangular support into account. For those macrostates we were able to prove that the resulting DFA implements the original Turing machine. In this sense, the restriction to uniform p.d.f.s with rectangular support introduces a contingent context from which symbolic computation emerges. (Note that uniform p.d.f.s also have maximal entropy).


\ack
This research was supported by a Heisenberg grant (GR 3711/1-1) of the German Research Foundation (DFG) awarded to PbG. Preliminary results have been presented at a special session ``Cognitive Architectures in Dynamical Field Theory'', that was partially funded by an EuCogIII grant, at the 2nd International Conference on Neural Field Theory, hosted by the University of Reading (UK). We thank Yulia Sandamirskaya, Slawomir Nasuto and Gregor Sch\"oner for inspiring discussions.



\end{document}